\pdfoutput=1
\documentclass[10pt, times, twocolumn]{article}
\usepackage{./latex8}
\usepackage{graphicx}
\usepackage{times}
\usepackage{setspace}

\newenvironment{tablehere}
  {\def\@captype{table}}
  {}

\makeatother

\begin{document}

\title{Structure and Interpretation of Computer Programs\thanks{In
reverence to the Wizard Book.}}

\author {
Ganesh Narayan, Gopinath K\\
    Computer Science and Automation\\
    Indian Institute of Science\\
    \{nganesh, gopi\}@csa.iisc.ernet.in\\
\and
Sridhar V\\
    Applied Research Group\\ Satyam Computers\\
    sridhar@satyam.com
}

\maketitle

\thispagestyle{empty}

\begin{abstract}
Call graphs depict the static, caller-callee relation between ``functions" in a
program. With most source/target languages supporting functions as the
primitive unit of composition, call graphs naturally form the fundamental
control flow representation available to understand/develop software. They are also the substrate on which various interprocedural analyses are
performed and are integral part of program comprehension/testing. Given
their universality and usefulness, it is imperative to ask if call graphs
exhibit any intrinsic graph theoretic features -- across versions, program
domains and source languages. This work is an attempt to answer these
questions: we present and investigate a set of meaningful graph measures that 
help us understand call graphs better; we establish how these measures
correlate, if any, across different languages and program domains; we also
assess the overall, language independent software quality by suitably
interpreting these measures.

\end{abstract}

\section{Introduction}


Complexity is one of the most pertinent characteristics of computer programs
and, thanks to Moore's law, computer programs are becoming ever larger and complex; it's not atypical for a software product to contain hundreds of
thousands, even millions of lines of code where individual components
interact in myriad of ways. In order to tackle such complexity, variety of
code organizing motifs were proposed. Of these motifs, functions form the most
fundamental unit of source code: software is organized as set of functions --
of varying granularity and utility, with functions computing various results on
their arguments. Critical feature of this organizing principle is that
functions themselves can call other functions. This naturally leads to the
notion of function call graph where individual functions are nodes, with 
edges representing caller-callee relations; indegree depicts the number of
functions that could call the function and outdegree depicts the number of
functions that this function can call. Since no further restrictions are employed,
the caller-callee relation induces a generic graph structure, possibly with loops
and cycles.

In this work we study the topology of such (static) call graphs. Our present
understanding of call graphs is limited; we know: that call graphs are
directed and sparse; can have cycles and often do; are not strongly connected;
evolve over time and could exhibit preferential attachment of nodes and edges.
Apart from these basic understanding, we do not know much about the topology of
call graphs.

\section{Contributions}
\label{sec:contri}
In this paper we answer questions pertaining to topological properties of call
graphs by studying a representative set of open-source programs. In particular,
we ask following questions: What is the structure of call graphs? Are there any
consistent properties? Are some properties inherent to certain programming
language/problem class? In order to answer these questions, we investigate
set of meaningful metrics from plethora of graph
properties~\cite{graph:properties:2007}. Our specific contributions are:

\textbf{1)} We motivate and provide insights as to why certain call graph properties
are useful and how they could help us develop better and robust software. 
\textbf{2)} We compare graph structure induced by different language paradigms under
an eventual but structurally immediate structure -- call graphs. The authors are unaware of any study that systematically compare the call graphs of different languages; in particular, the ``call graph" structure of functional languages.
\textbf{3)} Our corpus, being varied and large, is far more statistically representative compared to the similar studies~(\cite{valverde:small-wrld03},
\cite{baxter:java},\cite{myers:sftnet}).
\textbf{4)} We, apart from confirming previous results in a rigorous manner,
also compute new metrics to capture finer aspects of graph structure.
\textbf{5)} As a side effect, we provide a potential means to assess software
quality, independent of the source language.

\pagebreak
Rest of the paper is organized as follows. We begin by justifying the utility
of our study and proceed to introduce relevant structural measures in section
\ref{sec:struct}. Section \ref{sec:corpus:method} discusses the corpus and
methodology. We then present our the measurements and interpretations (Section
\ref{sec:intrprt}). We conclude with section \ref{sec:related} and
\ref{sec:finale}.

\section{Motivation}
Call graphs define the set of permissible interactions and information flows and could influence software processes in non trivial ways. In order to give
the reader an intuitive understanding as to how graph topology could influence
software processes, we present following four scenarios where it does.

\textbf{Bug Propagation Dynamics}
\label{sec:sub:bug}
Consider how a bug in some function affects the rest of the software. Let \texttt{foo}
call \texttt{bar} and \texttt{bar} could return an incorrect
value because of a bug  in \texttt{bar}.
if \texttt{foo} is to incorporate this return value in its part of
computation, it is likely to compute wrong answer as well; that is, \texttt{bar}
has infected \texttt{foo}. Note that such an infection is
contagious and, in principle, \texttt{bar} can infect any arbitrary function $f_n$ as long as $f_n$ is connected to \texttt{bar}. Thus connectedness as graph property trivially translates
to infectability. Indeed, with appropriate notions of infection propagation and
immunization, one could understand bug expression as an epidemic process. It is well known that
graph topology could influence the stationary distribution of this process. In particular, the
critical infection rate -- the infection rate beyond which an infection is not containable -- is
highly network specific; in fact, certain networks are known to have zero
critical thresholds~\cite{sfree:epidemic02}. It pays to know if call graphs
are instances of such graphs.

\textbf{Software Testing:}
Different functions contribute differently to software stability. Certain functions
that, when buggy, are likely to render the system unusable. Such functions,
functions whose correctness is central to statistical correctness of the
software, are traditionally characterized by per-function attributes like indegree and size. Such simple measure(s), though useful, fail to
capture the transitive dependencies that could render even a not-so-well
connected function an Achilles heel. Having unambiguous metrics that measure a
node's importance helps making software testing more efficient. Centrality is such a
measure that gives a node's importance in a graph. Once relevant centrality
measures were assigned, one could expend relatively more time testing central functions.
Or, equally, test central functions and their called contexts for prevalent
error modes like interface nonconformity, context disparity and the
likes~(\cite{fault85}, \cite{fault06}). By considering node centralities, one
could bias the testing effort to achieve similar confidence levels without a
costlier uniform/random testing schedule; though most developers intuitively know the
importance of individual functions and devise elaborate test cases to stress
these functions accordingly, we believe such an idiosyncratic methodology could
be safely replaced by an informed and statistically tenable biasing based
on centralities. Centrality is also readily helpful in software impact
analysis.

\textbf{Software Comprehension:}
Understanding call graph structure helps us to construct tools that assist the
developers in comprehending software better. For instance, consider a tool that
magically extracts higher-level structures from program call graph by grouping
related, lower-level functions. Such a tool, for example, when run on a kernel
code base, would automatically decipher different logical subsystems, say,
networking, filesystem, memory management or scheduling. Devising such a tool
amounts to finding appropriate similarity metric(s) that partitions the graph so that nodes within a partition are ``more" similar compared to nodes
outside. Understandably, different notions of similarities
entail different groupings. Recent studies show how network structure controls
such grouping~\cite{cluster:ensemble-sfree} and how per node graph metrics can
be used to improve the developer-perceived clustering
validity~(\cite{wu-mining}, \cite{matsuo02clustering}).

\textbf{Inter Procedural Analysis}
Call graph topology could influences both precision and convergence of Inter
Procedural Analysis (IPA). When specializing individual procedures in a
program, procedures that have large indegree could end up being less optimal:
dataflow facts for these functions tend to be too conservative as they
are required to be consistent across a large number of call sites.
By specifically cloning nodes with large indegree and by distributing the
indegrees ``appropriately" between these clones, one could specialize
individual clones better. Also, number of iterations an iterative IPA
takes compute a fixed-point depends on the $max$(longest path length,
largest cycle). 

\section{Statistical Properties of Interest}
\label{sec:struct}
As with most nascent sciences, graph topology literature is strewn with
notions that are overlapping, correlated and misused gratuitously; for clarity,
we restrict ourselves to following structural notions. A note on usage: we
employ graphs and networks interchangeably; $G=(V, E)$, $\mid V \mid = n$ and
$\mid E \mid = m$; $(i, j)$ implies $i$ calls $j$; $d_i$ denotes the degree of
vertex i and $d_{ij}$ denotes the geodesic distance between $i$ and $j$; $N(i)$
denotes the immediate neighbours of i; graphs are directed and simple: for
every $(i_1,j_1)$ and  $(i_2, j_2)$ present, either $(i_1 \ne i_2)$ or $(j_1
\ne j_2)$ is true.

Graphs, in general, could be modeled as \textit{random}, \textit{small world},
\textit{power-law}, or \textit{scale rich}, each permitting different dynamics.

\textbf{Random graphs: } random graph model~\cite{erdos:renyi:random}, is
perhaps the simplest network model: undirected edges are added at random between a fixed number $n$
of vertices to create a network in which each of the $\frac{1}{2}n(n-1)$
possible edges is independently present with some probability $p$, and the
vertex degree distribution follows Poisson in the limit of large $n$.

\textbf{Small world graphs:}  exhibit high degree of clustering and have
 mean geodesic distance $\ell$ -- defined as, 
$ \ell^{-1} = \frac{1}{n(n+1)}\sum_{i\ne j} d_{ij}^{-1} $ -- 
in the range of $\log n$; that is, number of vertices within a distance $r$ of a typical central vertex grows
exponentially with $r$~\cite{newman:struct}.

It should be noted that a large number of networks, including random networks,
have $\ell$ in the range of $\log n$ or, even, $\log \log n$. In this work, we
deem a network to be small world if $\ell$ grows sub logarithmically
and the network exhibits high clustering.

\textbf{Power law networks:} These are networks whose degree distribution follow
the discrete CDF: 
$P[ X > x ] \propto cx^{-\gamma}$ , 
where $c$ is a fixed
constant, and $\gamma$ is the scaling exponent. When plotted as a double
logarithmic plot, this CDF appears as a straight line of slope $-\gamma$.
The sole response of power-law distributions to conditioning is a change in
scale: for large values of $x$, $P[X > x | X > X_i]$ is identical to the
(unconditional) distribution $P[X > x]$. This ``scale invariance" of power-law
distributions is attributed as scale-freeness. Note that this notion of
scale-freeness does not depict the fractal-like self similarity in every scale.

Graphs with similar degree distributions differ widely in other structural
aspects; rest of the definitions introduce metrics that permit finer classifications.

\textbf{degree correlations:} In many real-world graphs, the probability of
attachment to the target vertex depends also on the degree of the source
vertex: many networks show assortative mixing on their degrees, that is, a
preference for high-degree nodes to attach to other high-degree node; others
show disassortative mixing where high-degree nodes consistently attach to low-degree ones.
Following measure, a variant of Pearson correlation coefficient~\cite{newman:assort02}, gives the degree correlation.
$\rho = \frac{m^{-1} \sum_i j_ik_i - [ m^{-1} \sum_i \frac{1}{2}(j_i +
k_i)]^2}{m^{-1} \sum_i \frac{1}{2}(j_i^2 + k_i^2) - [m^{-1} \sum_i
\frac{1}{2}(j_i + k_i)]^2}$ ,
where $j_i$, $k_i$ are the degrees of the vertices at the ends of $i^{th}$
edge, with $i = 1 \cdots m$. $\rho$ takes values in the range $ -1 \leq
\rho \leq 1$, with $\rho > 0$ signifying assortativity and $\rho < 0$
signifying dissortativity. $\rho=0$ when there is no discernible correlation
between degrees of nodes that share an edge.

\textbf{scale free metric:} a useful measure capturing the fractal nature
of graphs is scale-free metric $s(g)$~\cite{li:srich}, defined as:
$s(g) = \sum_{ (i,j) \in E} d_id_j$ ,
along with its normalized variant $S(g) = \frac{s(g)}{s_{max}}$; $s_{max}$ is
the maximal $s(g)$ and is dictated by the type of network
understudy\footnote{For unrestricted graphs, $s_{max} = \sum_{i=1}^n (d_i/2).
d_i^2$.}. Rest of the paper will use the normalized variant.

$s(g)$ is maximal when nodes with similar degree connect to each
other~\cite{srich:rearrangement}; thus, $S(g)$ is close to one for networks that
are fractal like, where the connectivity, at all degrees, stays similar. On the
other hand, in networks where nodes repeatedly connect to dissimilar nodes,
$S(g)$ is close to zero. Networks that exhibit power-law, but have have a scale
free metric $S(g)$ close to zero are called \textit{scale rich}; power-law
networks whose $S(g)$ value is close to one are called \textit{scale-free}.
Measures $S(g)$ and $\rho$ are similar and are correlated; but they
employ different normalizations and are useful in discerning different
features~\cite{li:srich}.

\textbf{clustering coefficient: } is a measure of how clustered, or locally
structured, a graph is: it depicts how, on an average, interconnected each
node's neighbors are. Specifically, if node $v$ has $k_v$ \emph{immediate}
neighbors, then the clustering coefficient for that node, $C_v$, is the ratio
of number of edges present between its neighbours $E_v$ to the total possible
connections between $v$'s neighbours, that is, $k_v(k_v-1)/2$. The whole graph
clustering coefficient, $C$, is the average of $C_v$s: that is, $C = \left< C_v
\right>_v = \left< \frac{2E_v}{k_v(k_v-1)} \right>_v$.

\textbf{clustering profile: } $C$ has limited use when immediate connectivity
is sparse. In order to understand interconnection profile of transitively
connected neighbours, we use clustering profile~\cite{clust:prof}: $C^d_k =
\frac{\sum_{\{i |d_i = k\}} C^d(i)}{|\{ i|d_i = k \}|}$ , where $C^d(i) =
\frac{|\{(j,k); j,k \in N(i)|{d_{jk} \in {G(V \setminus i)}} = d\}|}{{|N(i)|
\choose 2}}$. Note that, by this definition, clustering coefficient $C$ is
simply $C^1_k$,

\textbf{centrality:} of a node is a measure of relative importance of the node
within the graph; central nodes are both points of opportunities -- that
they can reach/influence most nodes in the graph, and of constraints -- that
any perturbation in them is likely to have greater impact in a graph. Many
centrality measures exist and have been successfully used in many
contexts~(\cite{pagerank:anatomy07}, \cite{puppin:java06}). Here we focus on
\textbf{betweenness centrality} $B_u$ (of node $u$), defined as the ratio of
number of geodesic paths that pass through the node ($u$) to that of the total
number of geodesic paths: that is, $B_u = \sum_{ij} \frac{\sigma (i, u,
j)}{\sigma(i,j)}$; nodes that occur on many shortest paths between other
vertices have higher betweenness than those that do not.

\textbf{connected components:} size and number of connected components gives us
the macroscopic connectivity of the graph. In particular, number and size
of strongly connected components gives us the extent of mutual recursion
present in the software. Number of weakly connected component gives us the
upper bound on amount of runtime indirection resolutions possible.

\textbf{edge reciprocity:} measures if the edges are reciprocal, that is, if
$(i,j) \in E$, is $(j, i)$ also $\in E$? A robust measure for reciprocity is
defined as~\cite{recipro:Garlaschelli}: $\rho  =
\frac{\varrho-\bar{a}}{1-\bar{a}}$
where $\varrho = \frac{\sum_{ij} a_{ij}a_{ji}}{m}$ and $\bar{a}$ is mean of
values in adjacency matrix. This measure is absolute: $\rho$ greater than zero
imply larger reciprocity than random networks and $\rho$ less than zero imply
smaller reciprocity than random networks.

\begin{figure}[t]
\centering
\includegraphics[height=1.5in, width=3.5in]{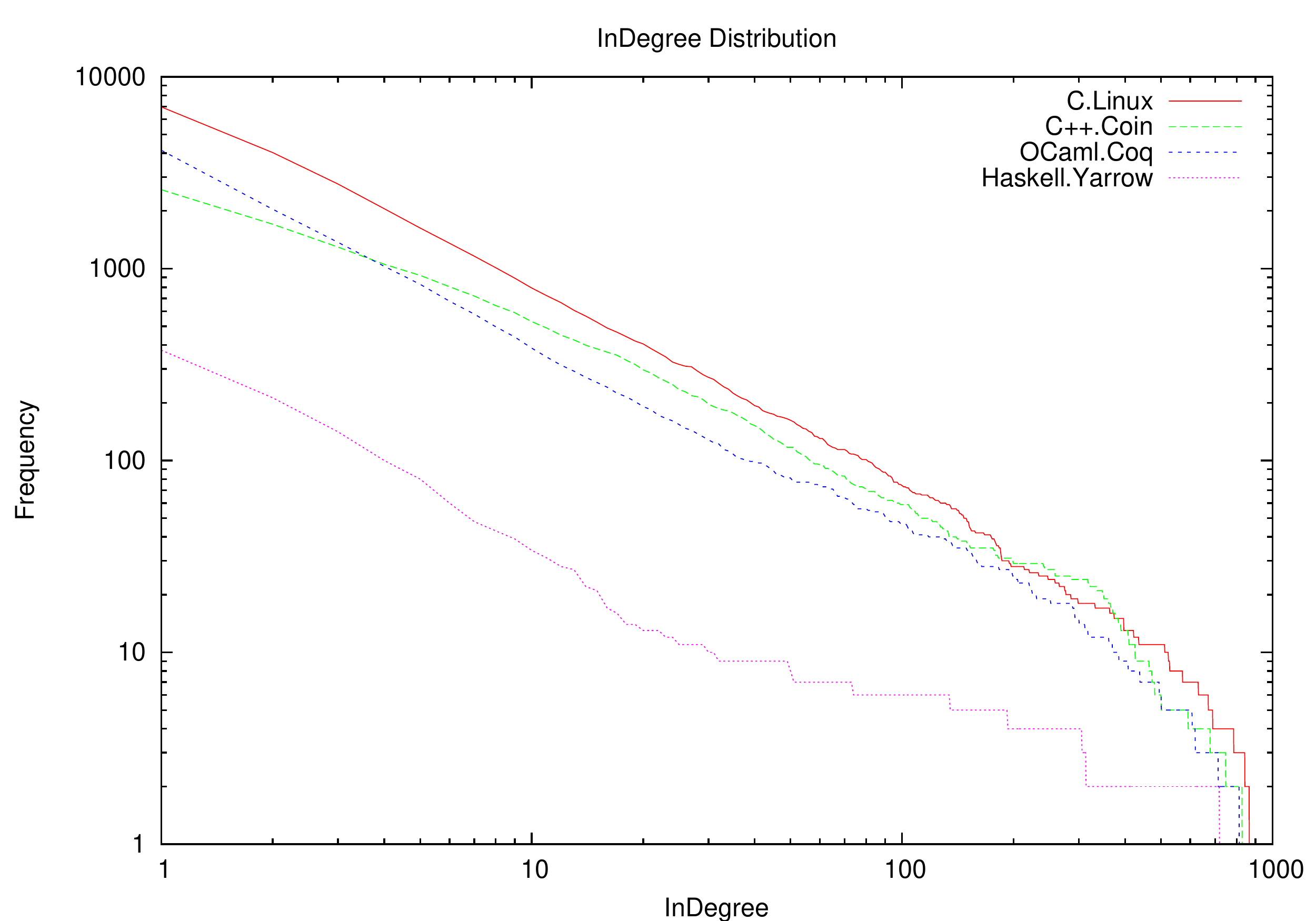}
\caption{Indegree Distribution}
\label{fig:in}
\end{figure}

\section{Corpora \& Methodology}
\label{sec:corpus:method}
We studied 35 open source projects. The projects are written in
four languages: C, C++, OCaml and Haskel. Appendix~\ref{app:1} enlists these
software, their source language, versiom,  domain and size: number of nodes $N$ and the
number of edges $M$. Most programs used are large, used by tens of thousands of users, written by hundreds of
developers and were developed over years. These programs are actively developed
and supported. Most of these programs -- from proof assistant to media player,
provide varied functionalities and have no apparent similarity or overlap in
usage/philosophy/developers; if any, they exhibit greater orthogonality: Emacs Vs
Vim, OCaml Vs GCC, Postgres Vs Framerd, to name a few. Many are stand-alone
programs while few, like glibc and ffmpeg, are provided as libraries. Some
programs, like Linux and glibc, have machine-dependent components while others
like yarrow and psilab are entirely architecture independent.

In essence, our sample is unbiased towards applications, source languages,
operating systems, program size, program features and developmental philosophy.
The corpus versions and age vary widely: some are few years old while others, like
gcc, Linux kernel and OCamlc, are more than a decade old. We believe that any
invariant we find in such a varied collection is likely universal.

We used a modified version of CodeViz~\cite{sft:codeviz} to extract call graphs
from C/C++ sources. For OCaml and Haskell, we compiled the sources to binary
and used this modified CodeViz to extract call graph from binaries. OCaml
programs were compiled using ocamlopt while for Haskell we used GHC. A note of
caution: to handle Haskell's laziness, GHC uses indirect jumps. Our tool,
presently, could handle such calls only marginally; we urge the reader to be
mindful of measures that are easily perturbed by edge additions.

We used custom developed graph analysis tools to measure most of the
properties; where possible we also used the graph-tool
software~\cite{sft:graphtool}. We used the largest weakly connected components
for our measurements. Component statistics were computed for the whole data
set.


\begin{figure}[t]
\includegraphics[height=1.5in, width=3.5in]{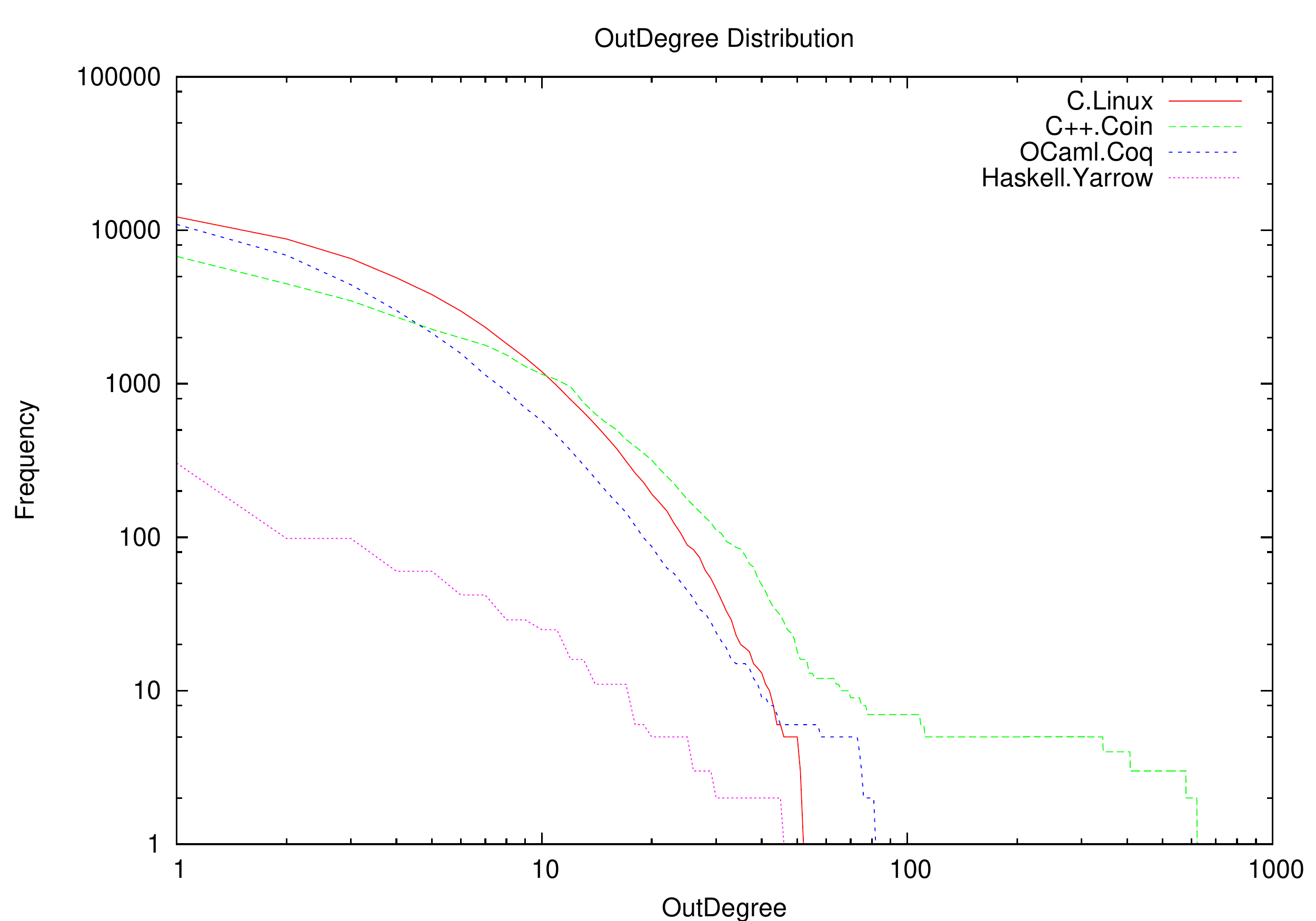}
\caption{Outdegree Distribution}
\label{fig:out}
\end{figure}

\section{Interpretation}
\label{sec:intrprt}
In the following section we walk through the results, discuss what these
results mean and why they are of interest to language and software communities.
Note that most plots have estimated sample variance as the confidence
indicator. Also, most graphs run a horizontal line that separates data from
different languages.

\textbf{Degree Distribution:} Fitting samples to a distribution is impossibly thorny: any sample is finite,
but of the distributions there are infinitely many. Despite the hardness of
this problem, many of the previous results were based either on visual
inspection of data or on linear regression, and are likely to be inaccurate~\cite{power:newman07}.

We use \textit{cumulative distribution} to fit the data and we compute the likelihood
measures for other distributions in order to improve the confidence using~\cite{power:newman07}. Figures
\ref{fig:in} and \ref{fig:out} depict how four programs written in four
different language paradigms compare; the indegree distribution permits
power-law ($2.3 \le \gamma \simeq  \le 2.9$) while the outdegree distribution
permits exponential distribution (Haskell results are coarse, but are valid). This observation, that
in and out degree distributions differ consistently across languages, is
expected as indegree and outdegree are conditioned very differently during the
developmental process.

\begin{figure}[t]
\centering
\includegraphics[height=1.5in, width=3.5in]{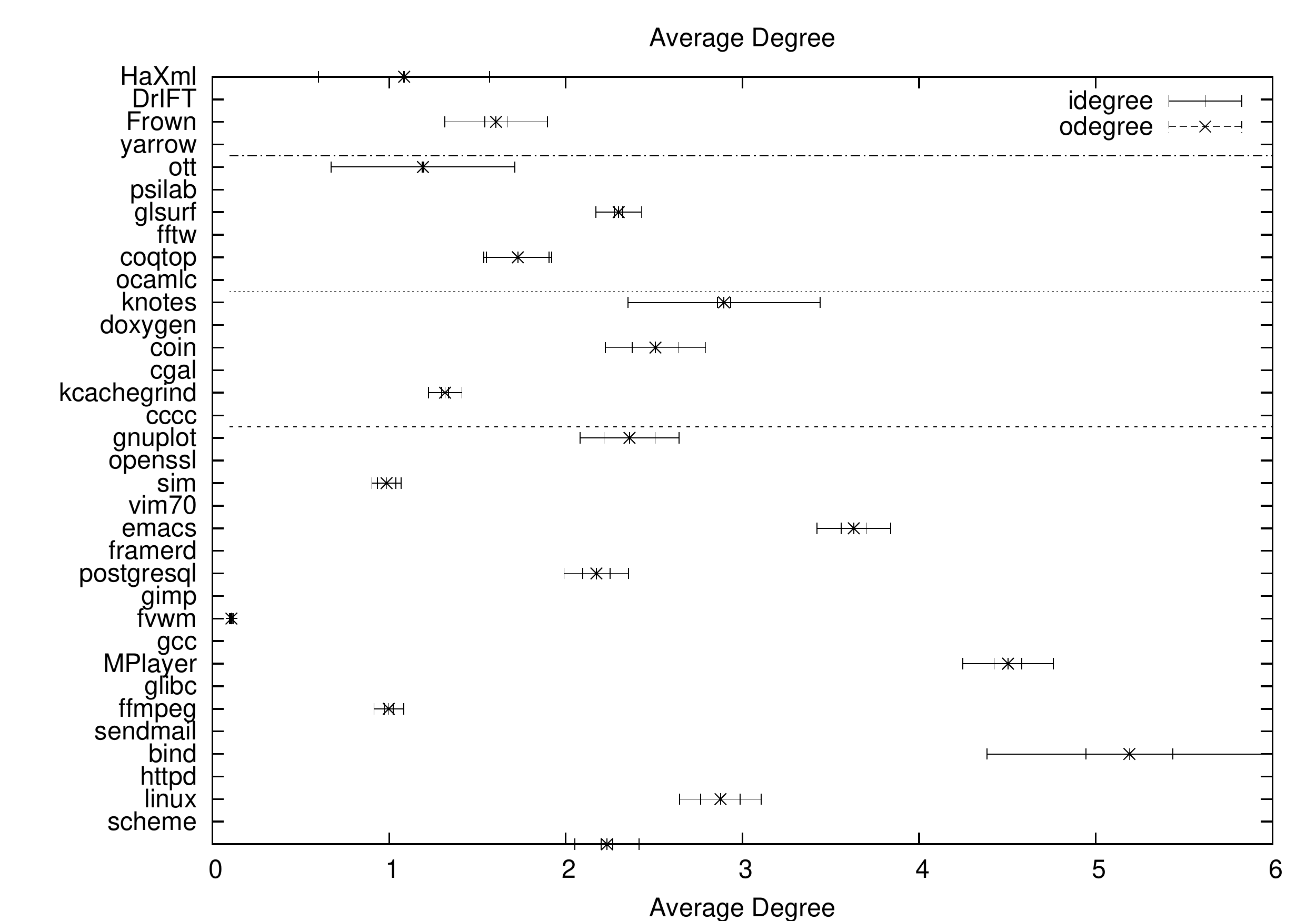}
\caption{Average Degree}
\label{fig:avg:deg}
\end{figure}

Outdegree has a strict budget; large, monolithic functions are difficult to
read and reuse. Thus outdegree is minimized on a local, immediate scale. On
the other hand, large indegree is implicitly encouraged, up to a point; indegree
selection, however, happens in a non-local scale, over a much larger
time period; usually backward compatibility permits lazy
pruning/modifying of such nodes. Consequently one would expect the variability
of outdegree -- as depicted by the length of the errorbar, to be far less
compared to that of the indegree. This is consistent with the observation (Fig.
\ref{fig:avg:deg}). Note that the tail of the outdegree is prominent in OCaml
and C++: languages that allow highly stylized call composition.

Such observations are critical as distributions portend the accuracy of sample
estimates. In particular, such distributions as power-law that permits
non-finite mean and variance -- consequently eluding central limit theorem, are
very poor candidates for simple sampling based analyses; understanding the
degree distribution is of both empirical and theoretical importance.

\begin{figure}[t]
\centering
\includegraphics[height=1.5in, width=3.5in]{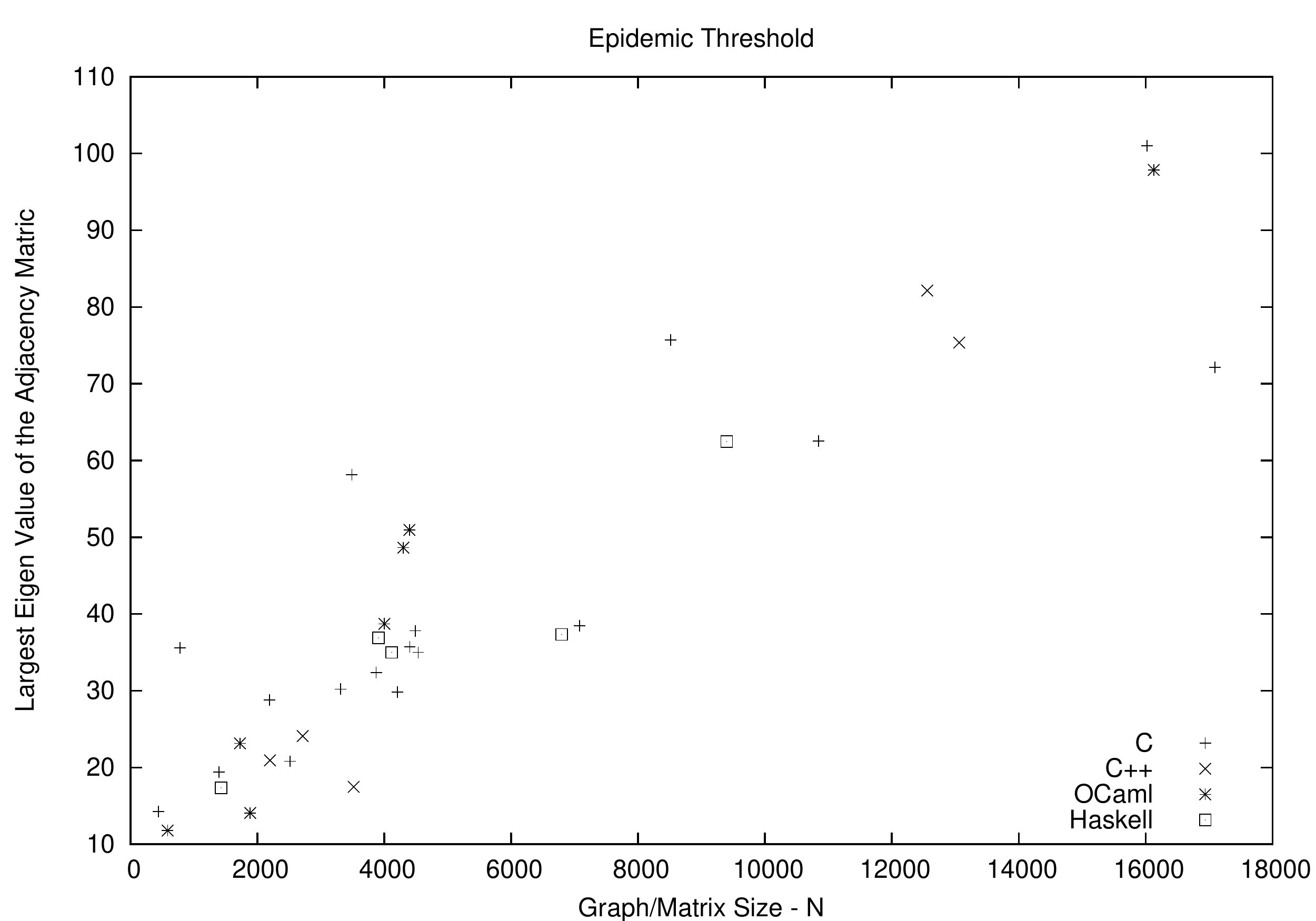}
\caption{Epidemic Threshold Vs $\mathbf{N}$}
\label{fig:epidemic:size}
\end{figure}

Consider the bug propagation process delineated in Section \ref{sec:sub:bug}. Assuming that the
inter-node bug propagation is Markovian, we could construct an
irreducible, aperiodic, finite state space  Markov chain (not unlike~\cite{pagerank:anatomy07}) with bug introduction rate $\beta$ and debugging
(immunization) rate $\delta$ as parameters. Note that this Markov chain has two
absorbing states: all-infected or all-cured. Equipped with these notions, we
could ask what is the minimal critical infection rate $\beta_c$ beyond which no
amount of immunization will help to save the software; below $\beta_c$ the
system exponentially converges to the good, all-cured absorbing state. It is
known that for a sufficiently large power-law network with exponent in the
range $2 < \gamma \le 3$, $\beta_{c}$ is zero~\cite{sfree:epidemic02}. Thus one is
tempted to conclude that, provided Markovian assumption holds, it is
statistically impossible to construct an all-reliable program. However that
would be inaccurate as the sum of indegree and outdegree
distribution\footnote{Bug propagation is symmetric: \texttt{foo} and \texttt{bar} can
pass/return bugs to one another.} indegree and outdegree need not follow
power-law. However a recent study~\cite{epidemic:spectrum:srds03} establishes
that, for finite networks, $\beta_c$ is bounded by the spectral diameter of the
graph; in particular, $\beta_c = \frac{1}{ \lambda_{1,A}}$, where
$\lambda_{1,A}$ is the largest eigenvalue of the adjacency matrix. Figure
\ref{fig:epidemic:size} depicts the relation between $\lambda_{1,A}$ and the
graph size, $n$. For a ``robust" software, we require $\beta_c$ to be large, or
equally, $\lambda_{1,A}$ to be small. However, it is evident from the plot that larger the graph, higher
the $\lambda_{1,A}$. This trend is observed uniformly across languages. Thus,
we are to conclude that large programs tend to be more fragile, confirming the
established wisdom. Another equally important inference one can make from the
indegree distribution is that uniform fault testing is bound to fail: should
one is to build a statistically robust software, testing efforts ought to be
heavily biased. These two inferences align closely with the common wisdom,
except that these inferences are rigorously established (and party explained)
using the statistical  nature of call graphs.

\begin{figure}[t]
\centering
\includegraphics[height=1.5in, width=3.5in]{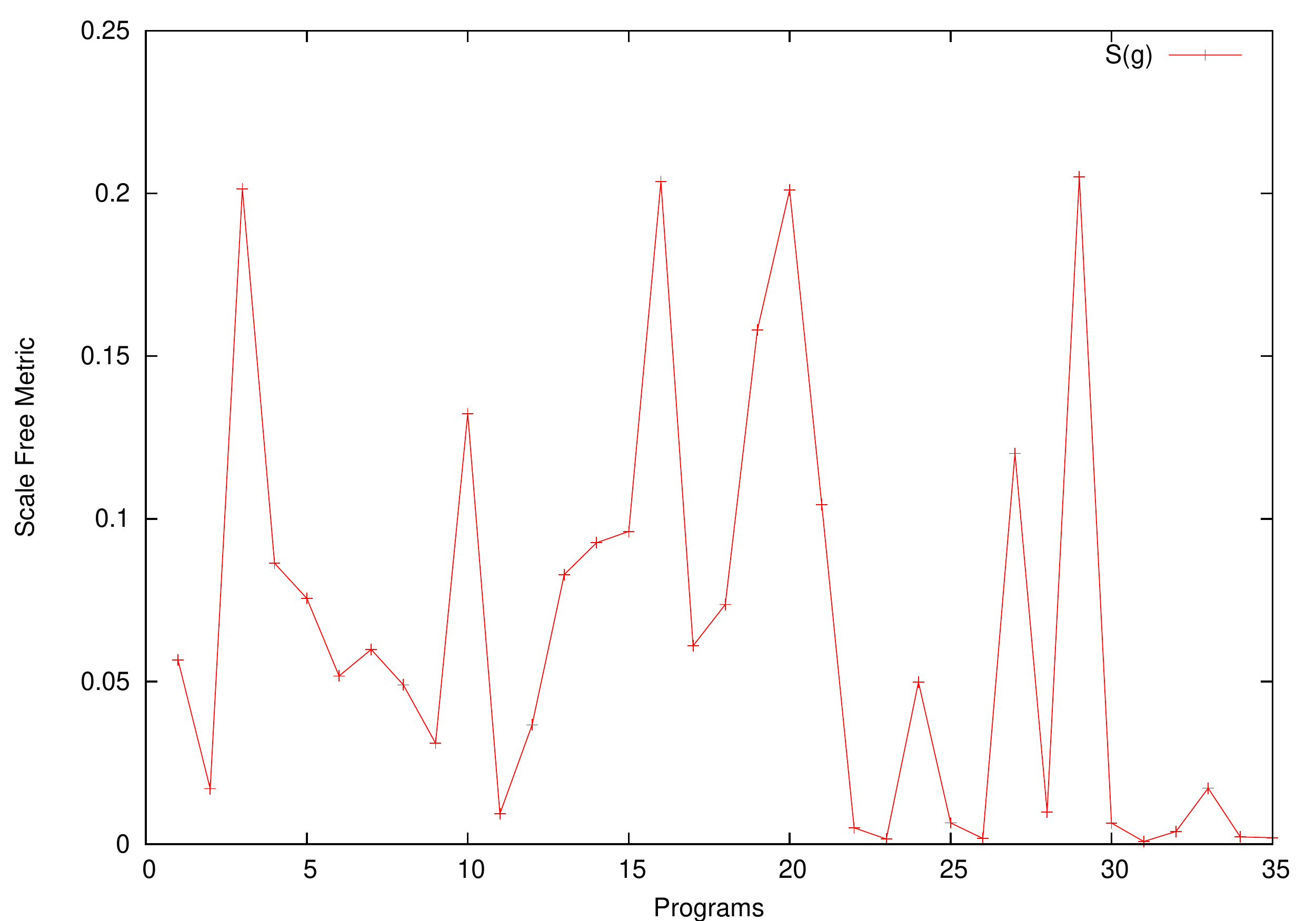}
\caption{Scale Free Metric}
\label{fig:sfree:metric}
\end{figure}

\textbf{Scale Free Metric:} Fig. \ref{fig:sfree:metric} shows how scale-free metric for symmetrized call
graphs vary with different programs. Two observations are critical: First,
$S(g)$ is close to zero. This implies call graphs are scale-rich and \emph{not}
scale-free. This is of importance because in a truly scale-free networks,
epidemics are even harder to handle; hubs are connected to hubs and the Markov
chain rapidly converged to the all-infected absorption state. In scale-rich
networks, as hubs tend to connect to lesser nodes, the rate of convergence is less
rapid. Second, $S(g)$ appears to be language independent\footnote{Except
Haskell; but this \emph{could} be an artifact of edge limited sample.}.
Both near zero and higher $S(g)$s appear in all languages. Thus call graphs,
though follow power-law for indegree, are not fractal like in the
self-similarity sense.

\begin{figure}
\centering
\includegraphics[height=1.5in, width=3.5in]{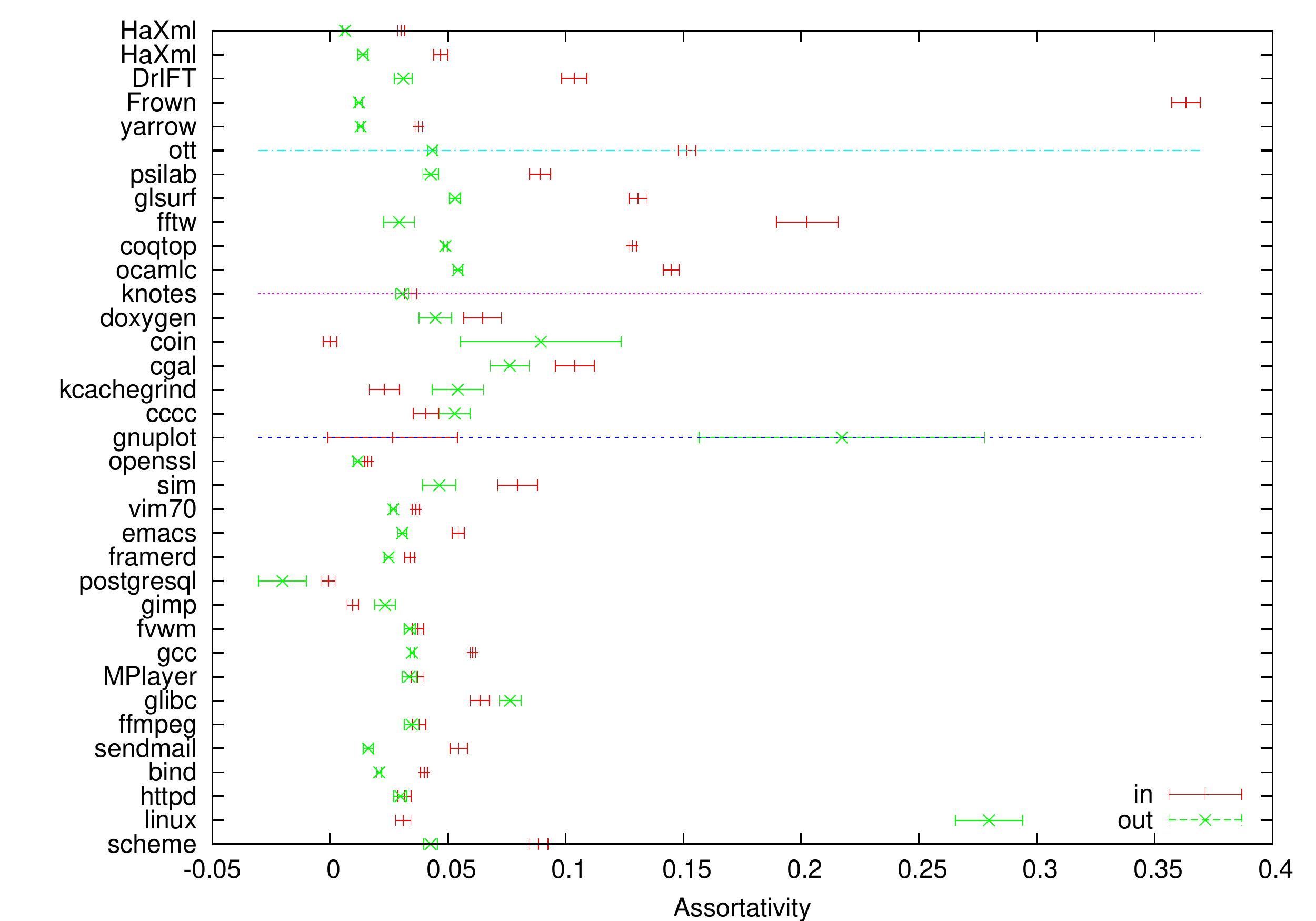}
\caption{Assortativity Coefficient}
\label{fig:assort}
\end{figure}

\textbf{Degree Correlation:} Fig \ref{fig:assort} show how input-input (i-i) and output-output (o-o) degrees
correlate with each other. These sets are weakly assortative, signifying
hierarchical organization.

But finer picture evolves as far as languages are concerned. C programs
appears to have very similar i-i and o-o profiles with o-o correlation being
smaller and comparable to i-i correlation. In addition, C's correlation measure
is consistently less than that of other languages and is close to zero; thus, C
programs exhibit as much i-i/o-o correlation as that of a random graph of similar size.
In other words, if \texttt{foo} calls \texttt{bar}, the number of calls
\texttt{bar} makes is
independent of the number of calls \texttt{foo} made; this implies less hierarchical
program structure as one would like the level $n$ functions to receive
fewer calls compared to level $n-1$ functions. For instance,
\texttt{variance(list)} is likely to receive fewer calls  compared to
\texttt{sum(list)}; we would also like level $n$ functions to have higher
outdegree compared to level $n-1$ functions. Thus, in a
highly hierarchical design, i-i and o-o correlations would be mildly assortative,
with i-i being more assortative. For C++, i-i and o-o differ and are not
ordered consistently. OCaml and Haskell exhibit marked difference in
correlations: as with C, the o-o correlation is close to zero; but, i-i
correlation is orders of magnitude higher than o-o correlation. That is, OCaml
forces nodes with ``proportional" indegree to pair up. If \texttt{foo} is has an
indegree X, \texttt{bar} is likely to receive, say, 2X indegree. One could interpret
this result as a sign of stricter hierarchical organization in functional
languages.

\begin{figure}
\centering
\includegraphics[angle=90,height=1.5in, width=3.5in]{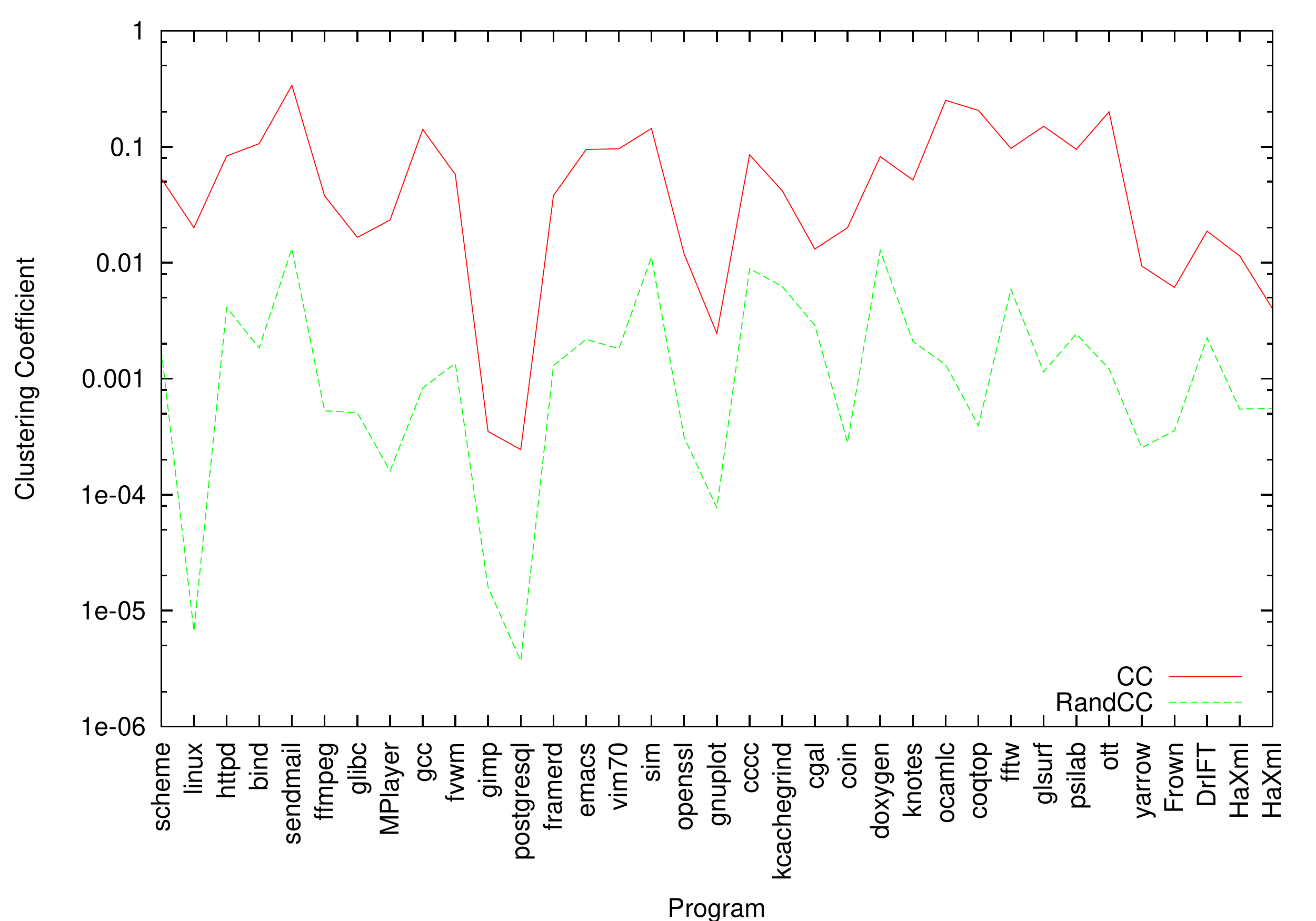}
\caption{Clustering Coefficient}
\label{fig:cc}
\end{figure}

\begin{figure}
\centering
\includegraphics[angle=90, height=1.5in, width=3.5in]{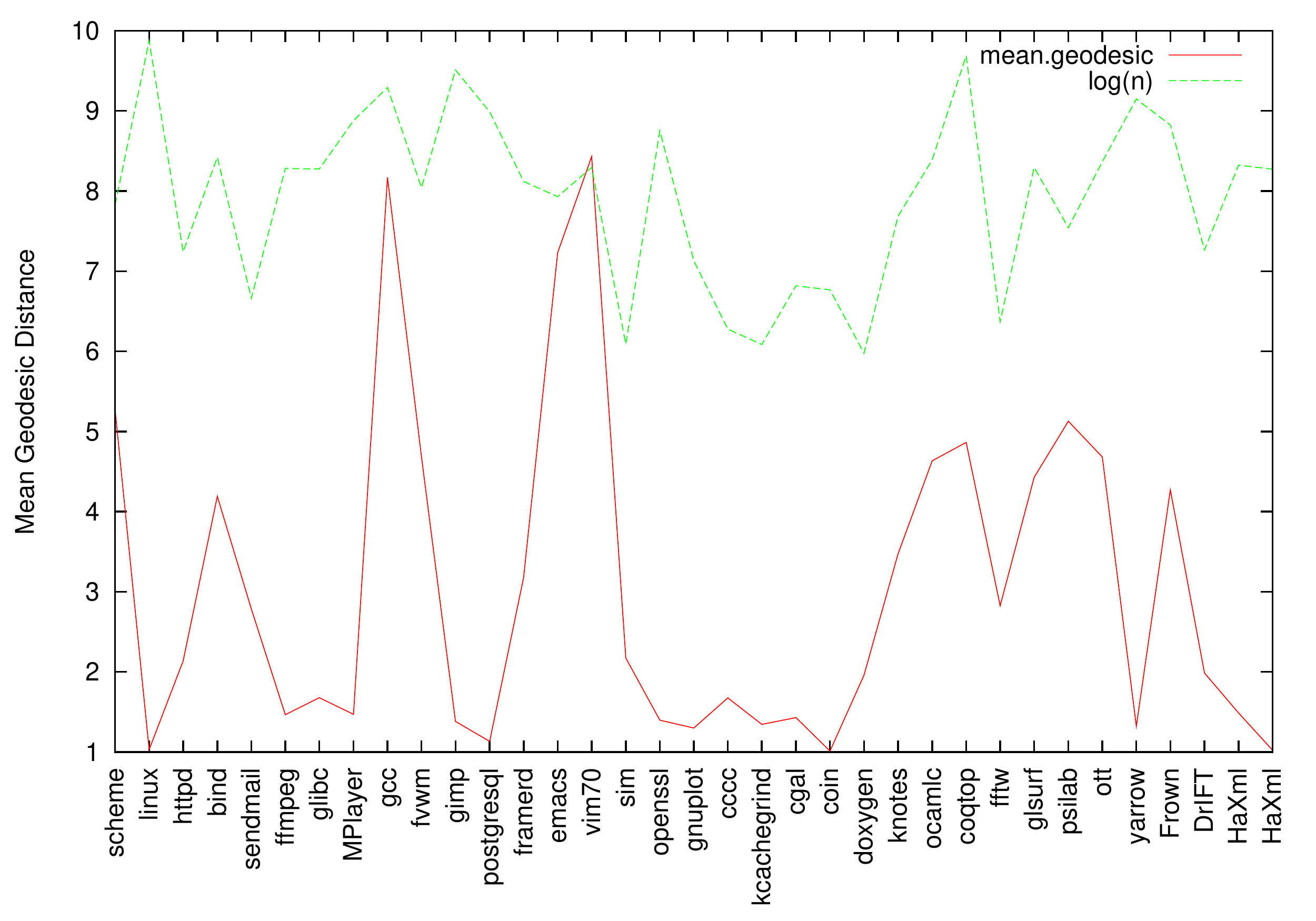}
\caption{Harmonic Geodesic Mean}
\label{fig:avggeo}
\end{figure}

\textbf{Clustering Coefficient:} Fig. \ref{fig:cc} depicts how the call graph clustering coefficients compare to
clustering coefficients of random networks of same size. Computed clustering
coefficients are orders of magnitude higher than their random counterpart
signifying higher degree of clustering. Also, observe that $\ell$, as depicted
is Fig. \ref{fig:avggeo}, is in the order of $\log n$. Together these
observations make call graphs decidedly small world, irrespective of the source
language.

We also have observed that average clustering coefficient for nodes of particular
degree, $C(d_i)$ follows power-law. That is, the plot of $d_i$ to $C(d_i)$
follows the power-law with $C(d_i) \propto d_i^{-1}$: high degree nodes exhibit
lesser clustering and lower degree notes exhibit higher clustering. It is also
observed that OCaml's fit for this power-law is the one that had least misfit.
Though we need further samples to confirm it, we believe functional languages
exhibit cleaner, non-interacting hierarchy compared to both procedural and OO
languages.

\begin{figure}
\centering
\includegraphics[angle=90,height=1.5in, width=3.5in]{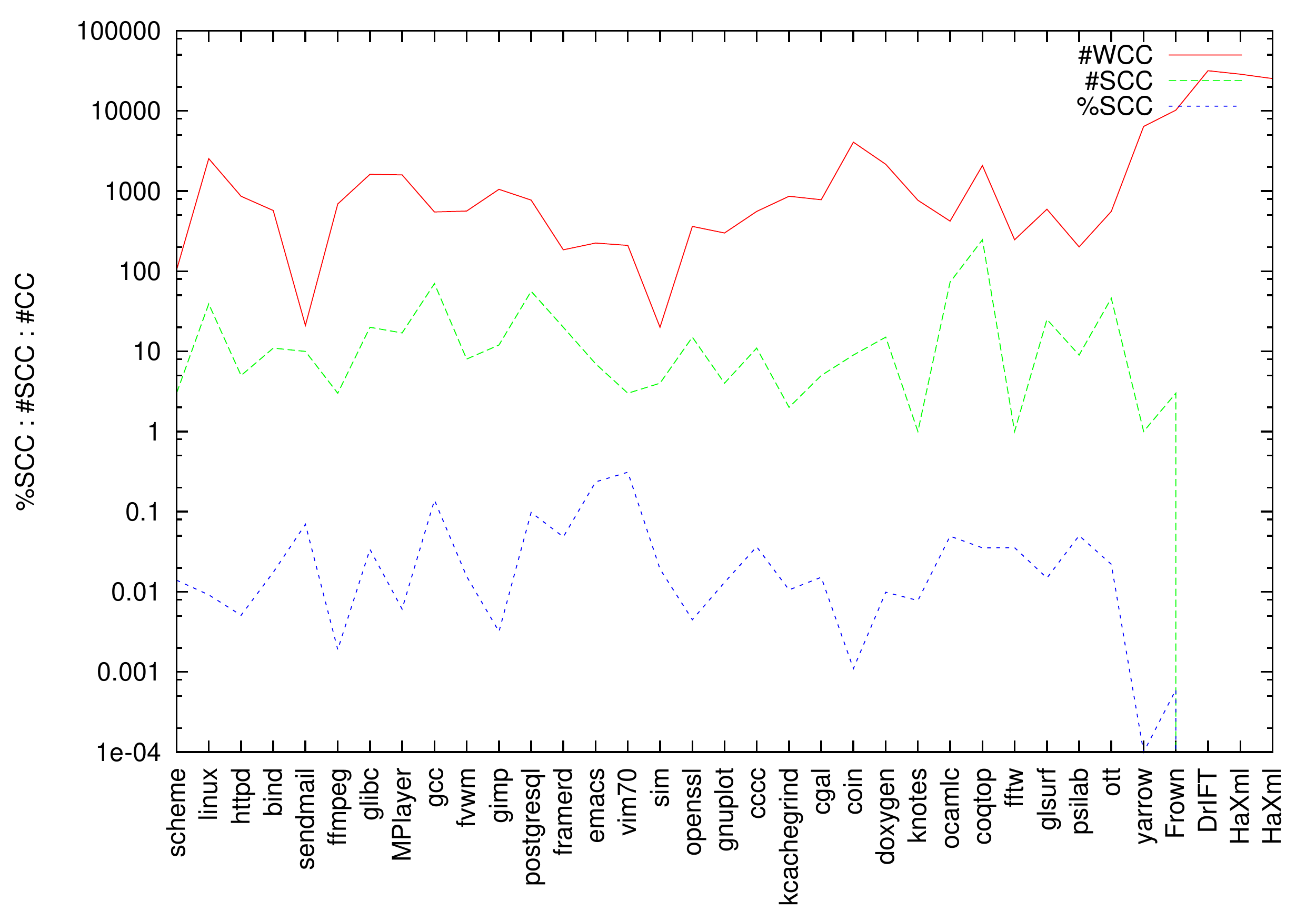}
\caption{Component Statistics}
\label{fig:comp}
\end{figure}

\textbf{Component Statistics:} Fig. \ref{fig:comp} gives us the components statistics for the data set. It
depicts the number of weakly connected components (\#WCC), number of strongly
connected components (\#SCC), and fraction of nodes in the largest strongly
connected component (\%SCC).

\#WCC is lower in C and OCaml. For C++ and Haskell, \#WCC is
higher compared to rest of the sample. This is an indication of lazy call
resolution, coinciding with the delayed/lazy bindings encouraged by both the
languages. The \#SCC values are highest for OCaml. This observation, combined
with reciprocity of OCaml programs, makes OCaml a language that encourages
recursion at varying granularity. On the other end, C++ rates least against
\#SCC values.

Another important aspect in Fig. \ref{fig:comp} is the observed values for
\%SCC; this fraction varies, surprisingly, from 1\% to 30\% of total number of
nodes. C leads the way with some applications, notably vim and Emacs, measuring
as much as 20 to 30\% for \%SCC. OCaml follows C with a moderate 2 to 6\% while
C++ measures  1\% to 3\%. We do not yet know why one third of an application
cluster to form a SCC. Also, \%SCC values say that certain languages, notably
OCaml, and programs domains (Editors: Vim and Emacs) exhibit significant mutual
connectivity.

\begin{figure}[t]
\includegraphics[height=1.5in, width=3.5in]{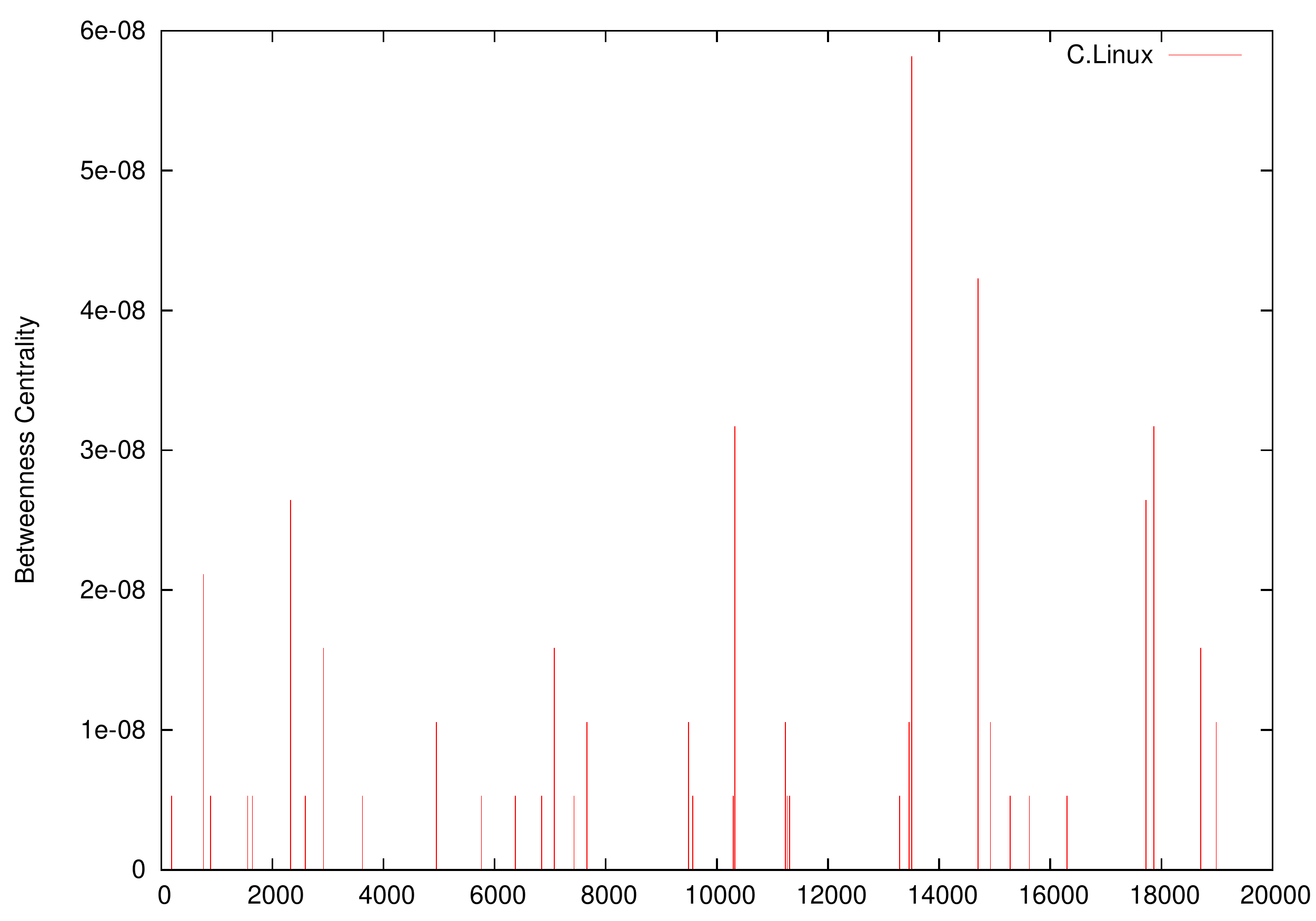}
\caption{Betweenness - C.Linux}
\label{fig:btw:c:lin}
\end{figure}

\begin{figure}
\centering
\includegraphics[angle=90,height=1.5in, width=3.5in]{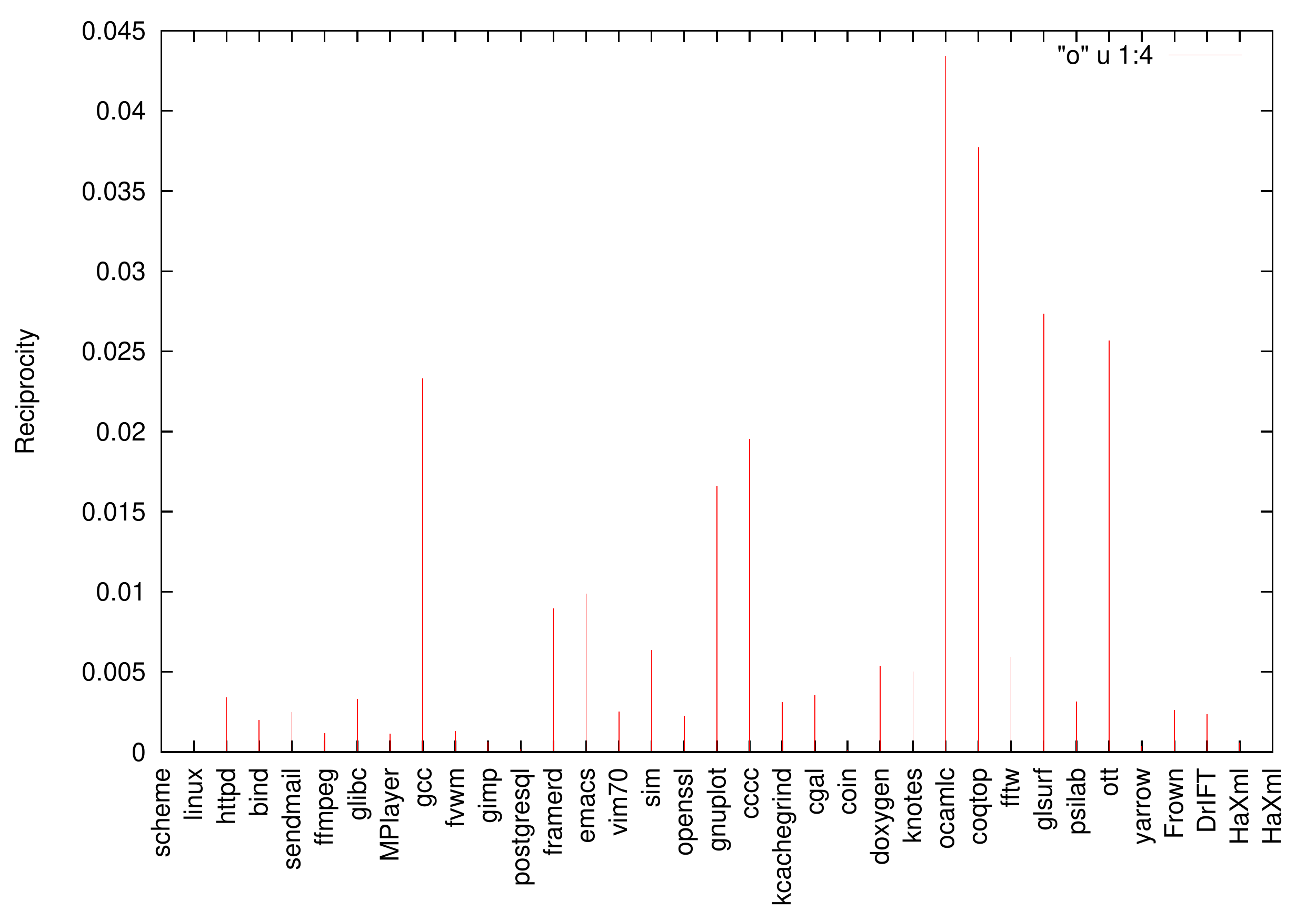}
\caption{Edge Reciprocity}
\label{fig:recipro}
\end{figure}

\textbf{Edge Reciprocity:} Fig. \ref{fig:recipro} shows the plot of edge reciprocity for various programs.
Edge reciprocity is a measure of \emph{direct} mutual recursion in the
software. High reciprocity in a layered system implies layering
inversion and we would, ideally, like a program to have negative reciprocity.

\pagebreak
Most programs exhibit close to zero reciprocity: most call graphs exhibit as
much reciprocity as that of random graphs of comparable size. None exhibit
negative reciprocity, implying no statistically significant preferential
selection to not to violate strict layering.

The software that had least reciprocity is the Linux kernel. Recursion of any
kind is abhorred inside kernel as kernel-stack is a limited resource; besides,
in a environment where multiple contexts/threads communicate using
shared memory, mutual recursion could happen through continuation flow, not
just as explicit control flow. Functional languages like OCaml naturally show
higher reciprocity. Another curious observation is that compilers, both OCamlc
and gcc, appear to have relatively higher reciprocity. This is the second
instance where applications (Compilers: GCC and OCamlc) determining the graph
property; this could be seen as a reflection of how compilers work: a great
deal of the lexing, parsing and semantic algorithms that compilers are based on
follow rich mutually recursive mathematical definitions.

\begin{figure*}[bth]
\includegraphics[angle=90, height=2in, width=7in]{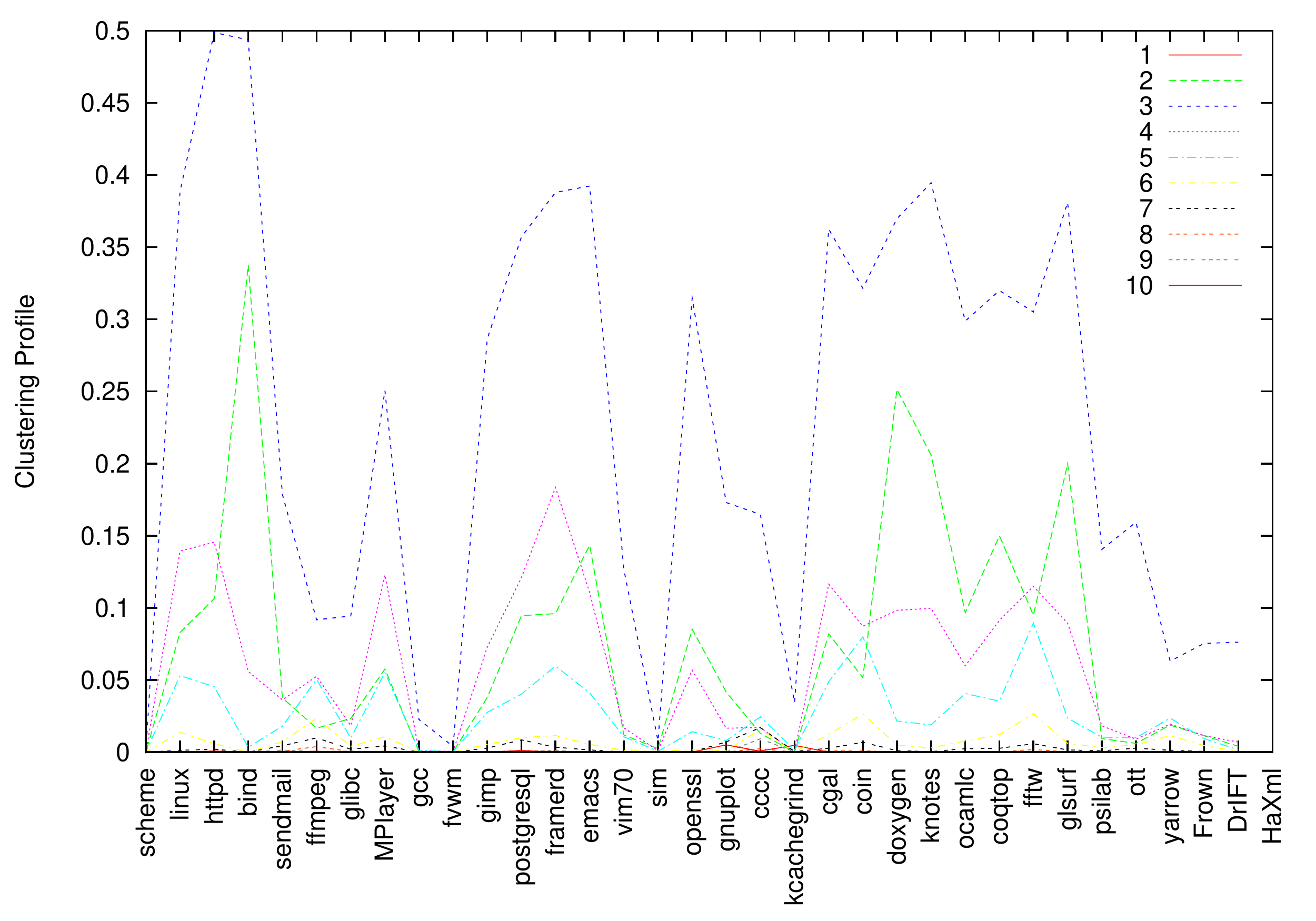}
\caption{Clustering Profile for Neighbours reachable in $\mathbf{k}$+2 hops}
\label{fig:ccprof}
\end{figure*}

\textbf{Clustering Profile:} As we see in Fig \ref{fig:ccprof}, Clustering Profile indeed gives us a better
insight. Y axis depicts the average clustering coefficient for nodes, say
$i$ and $j$, that are connected by geodesic distance $d_{ij}$.
In all the graphs observed, this average clustering increases up to $d_{ij}$=3
and falls rapidly as $d_{ij}$ increases further. We measured clustering
profile for degrees one to ten and the clustering profile appears to be
unimodal, reaching the maximum at $d_{ij}$=3, irrespective of language/program
domain. It suggests that maximal clustering occurs between nodes that are
separated \emph{exactly} by  five hops: clustering profile for a node $u$ is
measured with $u$ removed; so $d_{ij}$=3 is 5 hops in the original graph.
However exciting we find this result to be, we currently have no explanation
for this phenomenon.

\begin{figure}
\centering
\includegraphics[height=1.5in, width=3.5in]{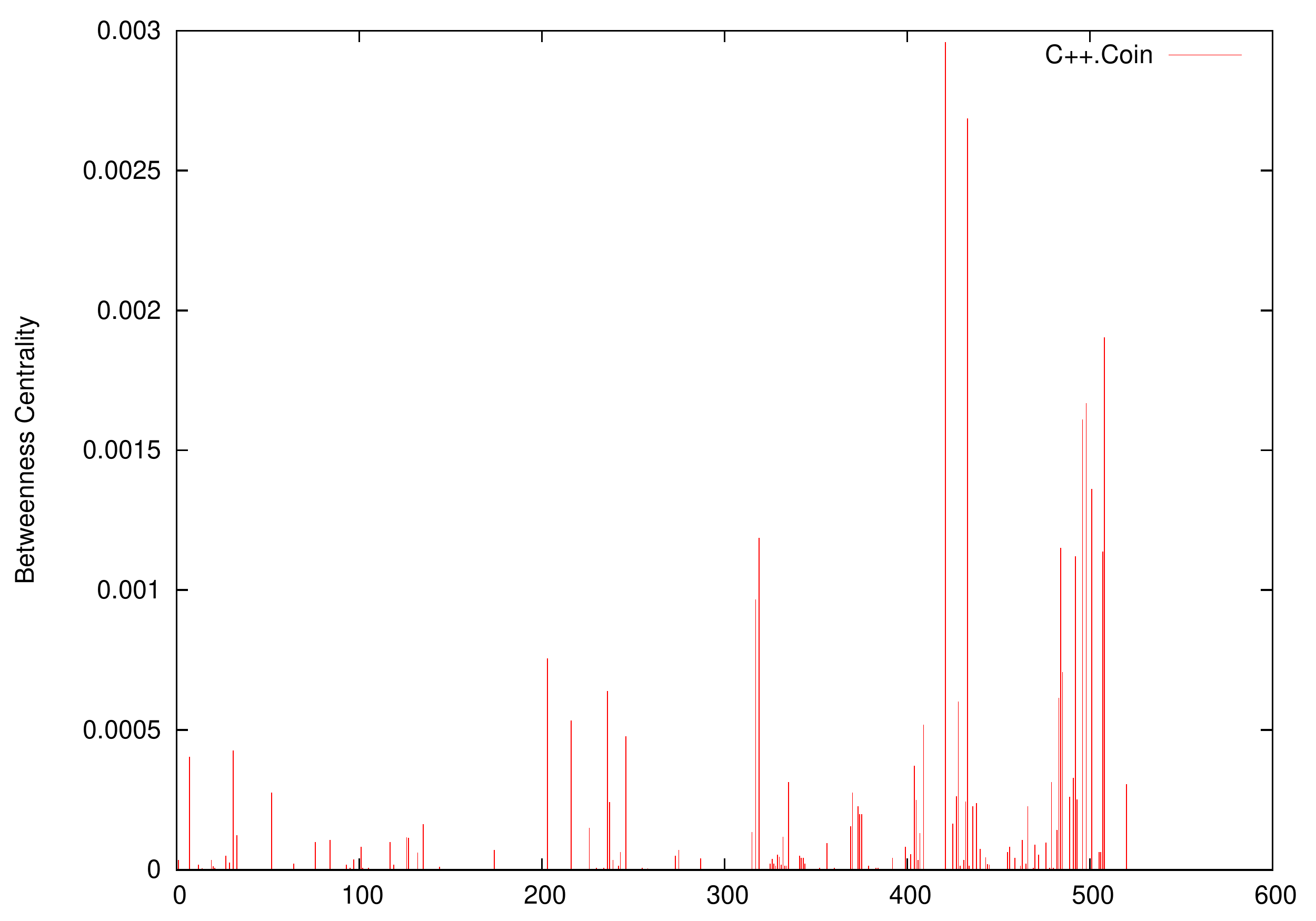}
\caption{Betweenness - C++.Coin}
\label{fig:btw:c++:coing}
\end{figure}
\begin{figure}
\includegraphics[height=1.5in, width=3.5in]{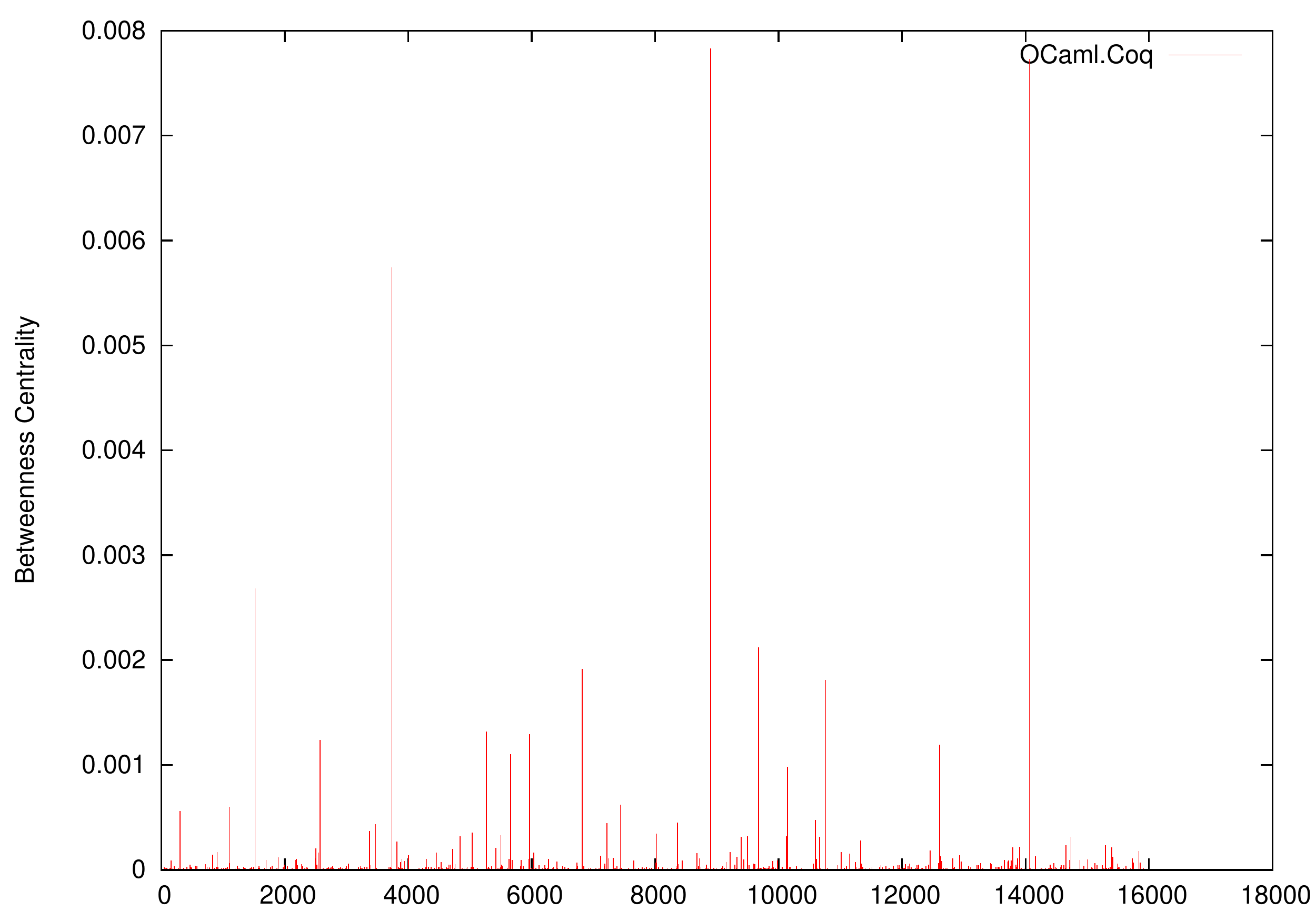}
\caption{Betweenness - OCaml.Coq}
\label{fig:btw:ocaml:coq}
\end{figure}

\textbf{Betweenness:} Fig. \ref{fig:btw:c:lin} to \ref{fig:btw:haskell:yarrow}
depict how betweenness centrality is distributed -- in different programs,
written in different different languages. Note that betweenness is not
distributed uniformly: it follows a rapidly decaying exponential distribution.
This confirms our observation that importance of functions is distributed
non-uniformly. Thus, by concentrating test efforts in functions that have
higher betweenness -- functions that are central to most paths -- we could test
the software better, possibly with less effort. An interesting line of
investigation is to measure the correlation between various centrality measures
and actual per function bug density in a real-world software.

\section{Related Work}
\label{sec:related}
Understanding graph structures originating from various fields is an active
field of research with vast literature; there is a renewed enthusiasm in
studying graph structure of software and many studies, alongside ours, report
that software graphs exhibit small-world and power-law properties.

\cite{myers:sftnet} studies the call graphs and reports that both
indegree and outdegree distributions follow power-law distributions and the
graph exhibits hierarchical clustering. But~\cite{valverde05:growth} suggests
that indegree alone follows power-law while the outdegree admits exponential
distribution. \cite{valverde05:growth} also suggests a growing network model
with copying, as proposed in~\cite{gnc:krapivsky05}, would consistently explain
the observations.

More recently, \cite{baxter:java} studies the degree distributions of various
meaningful relationships in a Java software. Many relationships admit power-law
distributions in their indegree and exponential distribution in their
out-degree. \cite{oo:geom} studies the dynamic, points-to graph of objects in
Java programs and found them to follow power-law.

Note that most work, excepting~\cite{baxter:java}, do not rigorously
compare the likelihood of other distributions to explain the same data. Power-law
is notoriously difficult to fit and even if power-law is a genuine fit,
it might not be the best fit~\cite{power:newman07}.

\section{Conclusion \& Future Work}
\label{sec:finale}
We have studied the structural properties of large software systems written in
different languages, serving different purposes. We measured various
finer aspects of these large systems in sufficient detail and have argued why
such measures could be useful; we also depicted situations where
such measurements are practically beneficial. We believe our study is a step
towards understanding software as an evolving graph system with distinct
characteristics, a viewpoint we think is of importance in developing and
maintaining large software systems.

There is lot that needs to be done. First, we need to measure the correlation
between these precise quantities and the qualitative, rule of thumb
understanding that developers usually possess. This helps us making such
qualitative, albeit useful, observations rigorous. Second, we need to verify
our finding over a much larger set to improve the inference confidence.
Finally, graphs are extremely useful objects that are analysed in a variety of
ways, each exposing relevant features; of these variants, the
authors find two fields very promising:
topological and  algebraic graph theories. In particular, studying call graphs
using a variant of Atkin's~{A-Homotopy theory is likely to yield interesting
results~\cite{atkin:graph}. Also, spectral methods applied to call graphs is
an area that we think is worth investigating.

\begin{figure}[t]
\includegraphics[height=1.5in, width=3.5in]{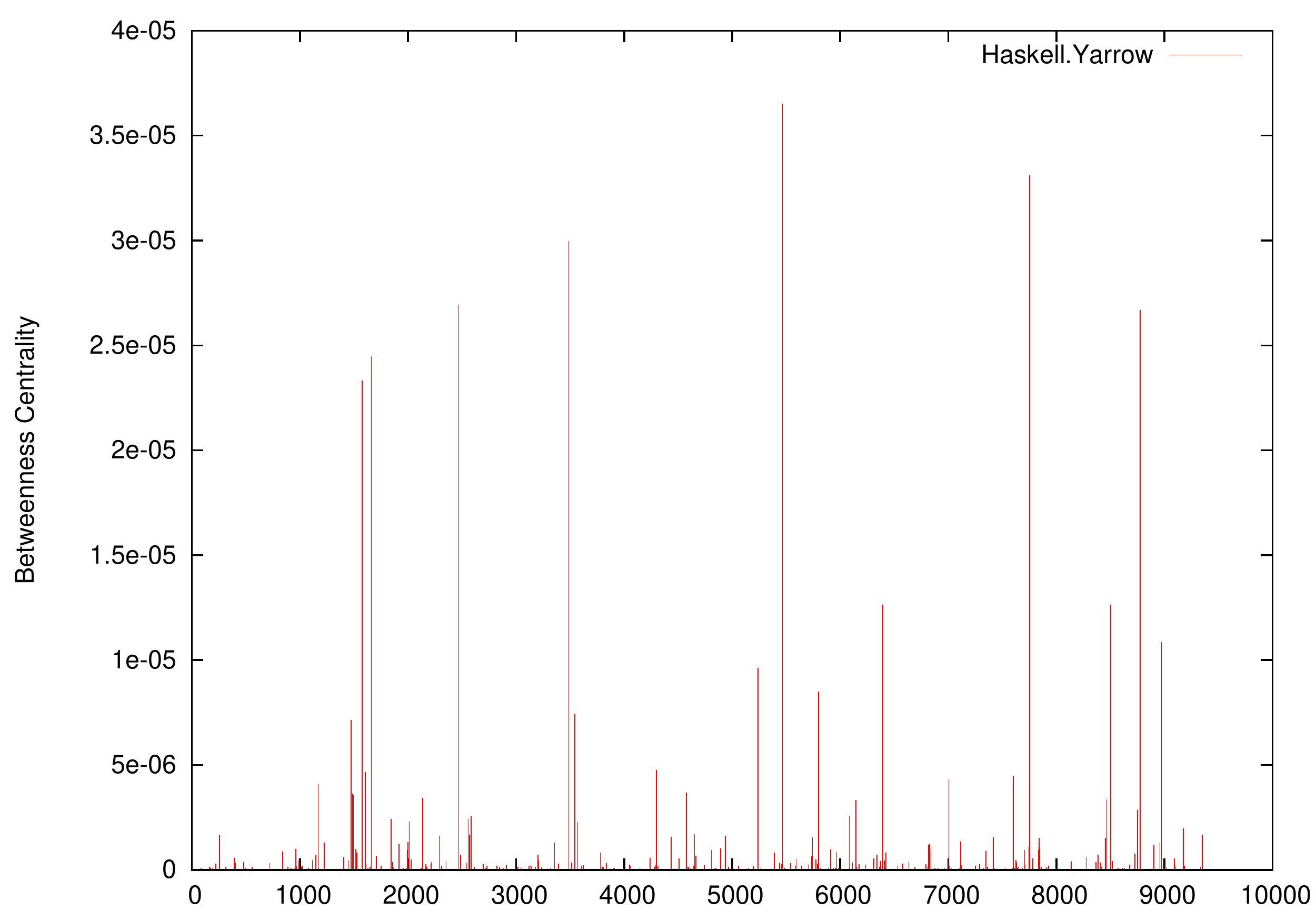}
\caption{Betweenness - Haskell.Yarrow}
\label{fig:btw:haskell:yarrow}
\end{figure}


\bibliographystyle{./latex8}
\bibliography{ref}

\pagebreak

\section*{Appendix I}
\label{app:1}
\begin{tablehere}
\center
\begin{tabular}{|c|l|l|l|c|c|}
\hline

\bf{*} & \bf{Lang} & \bf{Name.Version} & \bf{Appln. Domain} &  $\mathbf{N}$
& $\mathbf{M}$\\

\hline
\hline

1 & C &scheme.mit.7.7.1 & Interpreter & 2512 & 5610 \\ 
\hline
2 & C &linux.2.6.12.rc2 &  Kernel & 20165 & 70010 \\ 
\hline
3 & C &httpd.2.2.4 & Web Sever & 1396 & 4014 \\ 
\hline
4 & C &bind.9.4.1 & Name Server & 4534 & 18874 \\ 
\hline
5 & C &sendmail.8.12.8 & Mail Server & 783 & 4064 \\ 
\hline
6 & C &ffmpeg.2007.05 & Media Codecs & 4207 & 11692 \\ 
\hline
7 & C &glibc.2.3.6 &  C Lib & 4401 & 13972 \\ 
\hline
8 & C &MPlayer.1.0rc1 & Media Player & 7985 & 21744 \\ 
\hline
9 & C &gcc.4.0.0 & C Compiler  & 10848 & 48847 \\ 
\hline
10 & C &fvwm.2.5.18 & Win Manager & 3312 & 12052 \\ 
\hline
11 & C &gimp.2.3.9 & Image Editor & 16021 & 88473 \\ 
\hline
12 & C &postgresql.8.2.3 & R-DBMS  & 8517 & 41189 \\ 
\hline
13 & C &framerd.2.6.1 & OO-DMBS & 3490 & 17048 \\ 
\hline
14 & C &emacs.21.4 & ``Editor"  & 3872 & 13154 \\ 
\hline
15 & C &vim70 & Editor  & 4489 & 18368 \\ 
\hline
16 & C &sim.outorder.3.0 & $\mu$arch/ISA Sim & 442 & 1089 \\ 
\hline
17 & C &openssl.0.9.8e & Crypto Lib & 7078 & 21827 \\ 
\hline
18 & C &gnuplot.4.2.2 & Graph Plotting & 2191 & 7045 \\ 
\hline
19 & C++ & cccc.3.1.4 & Code Metrics & 1654 & 5627 \\ 
\hline
20 & C++ &kcachegrind.0.4 & Cache Analyser & 2593 & 8054 \\ 
\hline
21 & C++ &cgal.3.3 & CompGeom Lib  & 3151 & 7690 \\ 
\hline
22 & C++ &coin.2.4.6 & OpenGL 3D Lib  & 12963 & 51877 \\ 
\hline
23 & C++ &doxygen.1.5.3 & Doc Generator & 11723 & 31889 \\ 
\hline
24 & C++ &knotes.3.3 & PIM & 2174 & 4942 \\ 
\hline
25 & OCaml &ocamlc.opt.3.09 & OCaml Compiler & 4397 & 12732 \\ 
\hline
26 & OCaml &coqtop.opt.8 & Theorem Prover & 16126 & 51092 \\ 
\hline
27 & OCaml &fftw.3.2alpha2 & FFT Computing & 585 & 1011 \\ 
\hline
28 & OCaml &glsurf.2.0 & OpenGL Surface   & 4003 & 9173 \\ 
\hline
29 & OCaml &psilab.2.0 & Numeric Envirn & 1888 & 4341 \\ 
\hline
30 & OCaml &ott.0.10.11 & PL/Calculi Tool & 4300 & 11193 \\ 
\hline
31 & Haskell &yarrow.1.2 & Theorem Prover & 9397 & 15199 \\ 
\hline
32 & Haskell &Frown.0.6.1 & Parser Gen & 6796 & 10218 \\ 
\hline
33 & Haskell &DrIFT.2.2.1 & Typed Preproc & 1428 & 3292 \\ 
\hline
34 & Haskell &HaXml.1.Validate & XML Validate & 4117 & 7624 \\ 
\hline
35 & Haskell &HaXml.1.Xtract & XML grep & 3909 & 5242 \\ 
\hline
\end{tabular}

\end{tablehere}

\end{document}